\documentclass[letter,twocolumn]{jpsj2} 
\usepackage{color}

\def\kagome{{kagom{\' e }}}

\def\beq{\begin{equation}}
\def\eeq{\end{equation}}
\def\beqa{\begin{eqnarray}}
\def\eeqa{\end{eqnarray}}
\def\beqp{\begin{equation}\left\{\begin{array}{ll}}
\def\eeqp{\end{array}\right.\end{equation}}
\def\bfg{\begin{figure}}
\def\efg{\end{figure}}

\title{Slow Relaxation of Spin Structure in Exotic Ferromagnetic Phase of Ising-like Heisenberg 
Kagom{\' e} Antiferromagnets}

\author{Shu Tanaka$^1$\thanks{E-mail address: shu-t@spin.phys.s.u-tokyo.ac.jp}
and
Seiji Miyashita$^{1,2}$\thanks{E-mail address: miya@spin.phys.s.u-tokyo.ac.jp}}

\inst{$^1$ Department of Physics, The University of Tokyo, 7-3-1 Hongo, Bunkyo-ku, Tokyo, 113-0033, Japan.\\
$^2$ CREST, JST, 4-1-8, Honcho Kawaguchi, Saitama, 332-0012, Japan.}

\abst{
In the corner-sharing lattice, magnetic frustration causes macroscopic degeneracy in the ground state, 
which prevents systems from ordering.
However, if the ensemble of the degenerate configuration has some global structure, 
the system can have a symmetry breaking phenomenon and thus posses a finite temperature phase transition. 
As a typical example of such cases, the magnetic phase transition of the 
Ising-like Heisenberg antiferromagnetic model on the \kagome lattice has been studied. 
There, a phase transition of the two-dimensional ferromagnetic Ising universality class 
occurs accompanying with the uniform spontaneous magnetization.
Because of the macroscopic degeneracy in the ordered phase, 
the system is found to show an entropy-driven ordering process, 
which is quantitatively characterized by the number of 
``weathervane loop''(WL). We investigate this novel type of slow relaxation in regularly frustrated system.
}

\kword{
Slow Relaxation,
Phase Transition,
Entropy Effect,
Classical Heisenberg Spin System,
Antiferromagnetic Kagome System
}

\begin{document}
\maketitle


In the past several decades, there has been increasing interest in spin structure of several frustrated spin system, where non-collinear spin configurations appear due to competition of interaction.
The structure and symmetry of the non-collinear structures cause various types of interesting ordering processes\cite{Yoshimori,Kawamura}. 
In particular, antiferromagnets on the so-called corner-sharing lattices, e.g., \kagome and pyrochlore lattices, the frustration causes macroscopic degeneracy even in continuous spin systems, and thus in contrast to the case of triangular lattice, magnetic ordering in XY and Heisenberg models is difficult in the \kagome lattice. However, for classical Heisenberg antiferromagnet on the \kagome lattice, Harris {\it et al.}\cite{Harris} estimated magnetic susceptibility by means of high temperature expansion up to $\beta ^8$, and concluded that $\sqrt{3} \times \sqrt{3}$ structure should be selected for XY and Heisenberg spin system.
Huse and Rutenberg\cite{Huse} and Chalker {\it et al.}\cite{Chalker} studied hidden order parameter at $T \to 0+$.
Moreover, the concept of order by disorder has been discussed to choose an ordered state in energetically degenerate states\cite{Henley,Chubukov,Sachdev,Reimers}.
Experimentally, various \kagome antiferromagnetic compounds have been studied, e.g., jarosite-based materials AM$_{3}$(OH)$_6$(SO$_4$)$_2$ (A$^+$ is a caution such as K$^+$, Rb$^+$, NH$_4 ^+$, Tl$^+$, Ag$^+$ and Na$^+$, {\it etc.} and M$^{3+}$ is a magnetic ion such as Fe$^{3+}$ or Cr$^{3+}$ {\it etc.}) \cite{Takano,Wills1,Grohol,Bartlett,Townsend,Morimoto}, SrCr$_{9x}$Ga$_{12-9x}$O$_{19}$\cite{Ohta,Broholm}, Rb$_2$M$_3$S$_4$ (M$^{2+}$ is a magnetic ion such as Ni$^{2+}$ , Co$^{2+}$ , Mn$^{2+}$ )\cite{Fukamachi}, Zn$_x$Cu$_{4-x}$(OH)$_6$Cl$_2$\cite{Shores}, Ba$_2$Sn$_2$Ga$_3$ZnCr$_7$O$_{22}$\cite{Hagemann} and ${}^3$He on the graphite sheet\cite{Graywall} {\it etc.}

Beside the isotropic Heisenberg model, Kuroda and Miyashita\cite{Kuroda} pointed out that a thermodynamical phase transition takes place in the Ising-like Heisenberg \kagome antiferromagnets.
The properties of this phase transition were also recently investigated in details by Bekhechi and Southern\cite{Bekhechi}.
There, the ground state magnetization has two-fold degeneracy, and a phase transition which breaks this symmetry occurs at finite temperature.
Because of the nature of antiferromagnets on corner sharing lattices, there still remain macroscopic degenerate states in the magnetically ordered state. 
This macroscopic degeneracy causes peculiar properties in the ordered state such as the localization of the spin wave, {\it etc}. and the low temperature magnetic state is called ``exotic ferromagnetic phase''. 
As to experimental realization of this phase transition, Maegawa {\it et al.} found two cusps of the temperature dependence of the susceptibility in \kagome system NH$_4$Fe$_3$(OH)$_6$(SO$_4$)$_2$\cite{Maegawa}, and they concluded that the successive phase transition may be caused by an easy axis type anisotropy in the Heisenberg \kagome antiferromagnet.

The slow relaxation has been studied in the spin-glasses and also diluted ferromagnets, where the randomness of the system plays an important role for the slow dynamics. 
Recently, however, slow relaxation processes have been observed also in regularly frustrated spin systems such as (H$_3$O)Fe$_{3}$(OH)$_6$(SO$_4$)$_2$\cite{Wills2} experimentally.
Generally we may expect that the frustration prevents ordering and should cause the fast dynamics.
However, we pointed out that there exists slow relaxation in an entropy-driven ordering process in macroscopically degenerate states (i.e., a kind of the order by disorder mechanism) in the Ising-like Heisenberg \kagome antiferromagnetic model\cite{Tanaka}. 
Bekhechi and Southern\cite{Bekhechi} also found slow dynamics and discussed it in the context of the glassy dynamics.
In this Letter, we investigate microscopic properties of this slow relaxation by studying structure change in macroscopically degenerate states which are characterized by the ``weathervane loop''. 
We find an Arrhenius type relaxation process in the magnetically ordered state. 
The relaxation time of this slow process is much longer than that of magnetization. 


We consider Ising-like Heisenberg antiferromagnetic model in the \kagome lattice. 
The Hamiltonian is given by
\begin{equation}
 \mathcal{H} = -J \sum_{\left\langle i,j \right\rangle} 
 \left( S_i^x S_j^x + S_i^y S_j^y + \Delta S_i^z S_j^z
 \right),\,\,\, \left( \Delta > 1 \right), 
\end{equation}
where $\left\langle i,j \right\rangle$ denotes nearest neighbor pairs
in the \kagome lattice and $\Delta$ denotes Ising-like anisotropy.

The \kagome lattice has a corner-sharing structure of triangle clusters and the spins on a triangle cluster consists of the set ($S_{\alpha}$,$S_{\beta}$,$S_{\gamma}$) in the ground state.
For the Ising-like Heisenberg antiferromagnetic model, these are given by 
$S_{\alpha} = \left( 0,0,1 \right)$, 
$S_{\beta} = \left( \sin\theta\cos\phi,\sin\theta\sin\phi,-\cos\theta\right)$ and 
$S_{\gamma} = \left( -\sin\theta\cos\phi,-\sin\theta\sin\phi,-\cos\theta\right)$, 
where $\cos\theta=\Delta/(1+\Delta)$. The angle $\phi$ is any angle denoting the freedom of rotation of $2\pi$.

There are two types of ground state configurations at each triangle.
One is the one mentioned above, which we call ``up-down-down'' configuration. The other is the one of up-side-down structure, that is $S_{\alpha} = \left( 0,0,-1 \right)$, {\it etc.}, which we call ``down-up-up'' configuration.
Although the positions of the spins ($S_{\alpha}$,$S_{\beta}$,$S_{\gamma}$) is not fixed, the ground state has two-fold degeneracy. That is, if one triagle has the ``up-down-down'' configuration, all others must have the same type.
Because ``up-down-down'' configuration has a nonzero negative magnetization, and the ``down-up-up'' configuration has the opposite one, this symmetry breaking causes appearance of the spontaneous magnetization. That is, all the triangles have the local magnetization $M=\pm(1-2\cos\theta)$, and the symmetry breaking of the magnetization takes place, which causes the thermodynamic phase transition at finite temperature where $Z_2$ symmetry is broken in spite of no sublattice long range order exists\cite{Kuroda}.

\begin{figure}[h]
\begin{center}
\includegraphics[height=3cm]{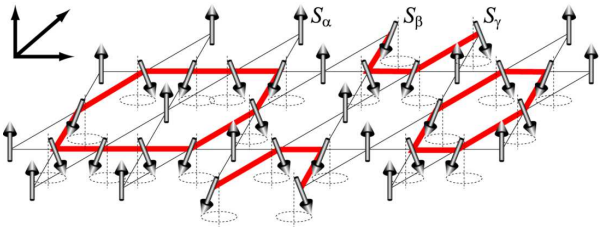}
\end{center}
\caption{(Color online) Spins on each triangle cluster consist of the three types ($S_\alpha$, $S_\beta$, $S_\gamma$).
The thick line denotes the WL.}
\label{Fig:kagomeWV}
\end{figure}

If we connect the spins of $S_{\beta}$ or $S_{\gamma}$ on the lattice, closed loops are formed as depicted by the bold red lines in Fig.~\ref{Fig:kagomeWV}.
These closed loops are called ``weathervane loop''(WL).
There are macroscopic ways to draw the WLs on the lattice.
Indeed the number of the ways is given by Udagawa {\it et al.} which is the same as that of the $S=1/2$ Ising antiferromagnetic \kagome model in the magnetic field\cite{Udagawa}. 
All of them are energetically degenerate.

In macroscopic energetically degenerate states the effect of spin configuration entropy sometimes plays an important role to determine the thermodynamical ordered state. 
In the present model, there exists the freedom of rotation $2\pi$ at each WL, which causes the residual entropy.
We investigate dynamics of the number of the WLs $n_{\mathrm{loop}}$ to characterize the relaxation of spin structure.

\begin{figure}[h]
\begin{center}
\includegraphics[height=5cm]{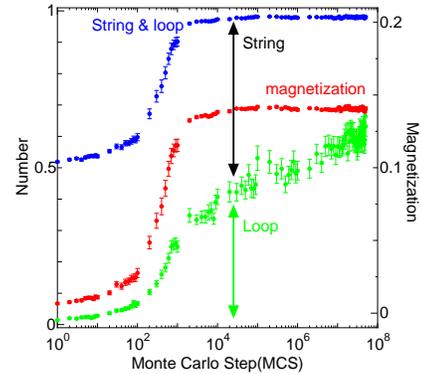}
\end{center}
\caption{(Color online) Relaxation of $n_{\mathrm{loop}}$, string and magnetization at $T=0.05J$ from random configuration.}
\label{Fig:loopdy}
\end{figure}

The maximum number of WLs $N_{\mathrm{loop}}^{\mathrm{max}}$, is realized in the 
$\sqrt{3} \times \sqrt{3}$ structure, in which all WLs are a simple hexagon which has minimum length.
In this case, $N_{\mathrm{loop}}^{\mathrm{max}}$ is $N/9$, where $N$ is the number of sites.
At finite temperature, there exist broken WLs with energetic defects.
We call these broken loops ``weathervane strings''.
We studied relaxation processes from a random configuration by the thermal bath Monte Carlo method 
in the lattice with $N=675$. 
In the present work, we adopted $\Delta = 3$.
We obtained the data by averaging over 24 independent runs.
We depict the dynamics of these quantities at $T=0.05J$ 
which is below $T_{\mathrm{c}} \left( =0.076J \pm 0.001J \right)$\cite{Bekhechi}
in Fig.~\ref{Fig:loopdy}.

When the energy or the magnetization (red points in Fig.~\ref{Fig:loopdy}) reaches to its equilibrium value, the total length of the WLs and strings 
(blue points in Fig.~\ref{Fig:loopdy}) also reaches to the equilibrium value.
Here the magnetization means the average of the $z$-component of spins.
However, the ratio of the contribution of the loops (green points in Fig.~\ref{Fig:loopdy}) increases in time as depicted in the Fig.~\ref{Fig:loopdy}.

\begin{figure}[h]
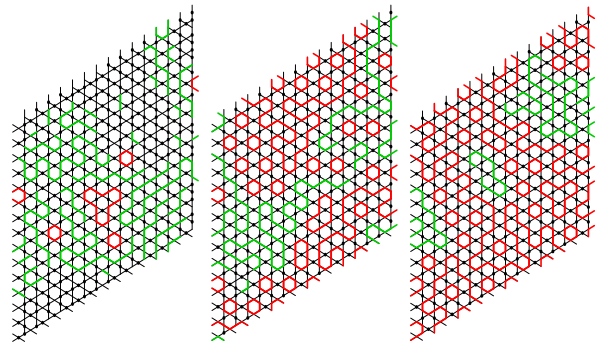

\begin{center}
\includegraphics[height=4.5cm]{15540Fig3a.ps}
\includegraphics[height=4.5cm]{15540Fig3b.ps}
\includegraphics[height=4.5cm]{15540Fig3c.ps}
\end{center}
\caption{(Color online) Snapshots at $10^2$MCS, $10^7$MCS, $4\times 10^7$MCS. The bold lines denote WLs and strings, respectively.}
\label{Fig:snapshot}
\end{figure}

Figures~\ref{Fig:snapshot} are the snapshots of spin configuration at $10^2$, $10^7$, and $4\times 10^7$ MCS. 
The red closed and green open lines denote WLs and weathervane strings, respectively.

Let us study temperature dependence of the entropy of a WL of the length $L$ at finite temperatures.
At low temperatures, the low energy excitation is restricted in the freedom of the angle $\phi$ of $\{ S_\beta \}$ and $\{ S_\gamma \}$.
This degree of freedom is expressed by the antiferromagnetic XY chain with periodic boundary condition.
$\mathcal{H}=-J\sum_{\left\langle i,j \right\rangle} \mathbf{S}_i \cdot \mathbf{S}_j$,
where $\mathbf{S}_i = \left( \cos \phi, \sin \phi\right)$.
The entropy per spin decreases with the length and the effect of spin chain length becomes more significant when the temperature decreases.
Because the entropy per spin decreases monotonously with the length of WL, the short loops are entropically preferable, and thus $n_{\mathrm{loop}}$ should increase at low temperatures.
What is more, at $T \to 0+$, we expect that $n_{\mathrm{loop}}$ must be the maximum value 
and the spin structure becomes the so-called $\sqrt{3} \times \sqrt{3}$ structure.



Next, we study the relaxation of magnetization and $n_{\mathrm{loop}}$. We ready the three types of initial configurations, {\it i.e.} 
(a) the $\sqrt{3} \times \sqrt{3}$ structure,
(b) the $q=0$ structure and 
(c) a random structure.
The configurations (a) and (b) are typical ground states of the present model, and the configuration (c) corresponds to a state just after quench the temperature from a high temperature.
\begin{figure}[h]
\begin{center}
\includegraphics[height=3.7cm]{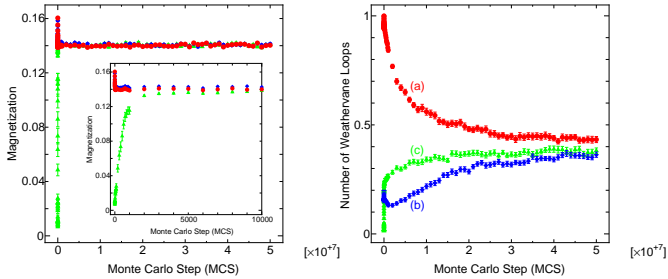}
\end{center}
\caption{(Color online) Relaxation of the magnetization and $n_{\mathrm{loop}}$ at $T=0.05J$ from 
(a) $\sqrt{3} \times \sqrt{3}$ configuration,
(b) $q=0$ configuration and
(c) random configuration.
}
\label{Fig:relaxmagloop}
\end{figure}

In Figs.~\ref{Fig:relaxmagloop}, the relaxation processes at $T=0.05J$ are plotted.
In the cases (a) and (b), the magnetization is maximum at $t=0$, and it relaxes very fast to uniformly magnetized ordered state. 
The relaxation of magnetization to the equilibrium is depicted in the inset.
In contrast, in the case (c), i.e., from a random state, it takes some time to realize the uniformly magnetized state. Thus we regard the relaxation time in the case (c) as the intrinsic relaxation time of the magnetization $\tau_{\mathrm{mag}}$. 

The relaxation of $n_{\mathrm{loop}}$ is much slower. In the case (a) $n_{\mathrm{loop}}$ starts with the maximum value, while in the case (c) it starts with zero. 
In all the cases, (a), (b) and (c), the numbers seem to reach the same saturated value. 
This observation indicates that there exists stable equilibrium state for $n_{\mathrm{loop}}$.
The relaxation time $\tau_{\mathrm{loop}}$ is also about the same, although in the case (b) a non-monotonic process is observed in the early state where the weathervane lines in the initial $q=0$ state are broken into short strips. Thus, we expect that there exists an intrinsic relaxation time for $n_{\mathrm{loop}}$.
From the figure we estimate that $\tau_{\mathrm{mag}}$ is about $10^4$ MCS, and $\tau_{\mathrm{loop}}$ is of the order $10^7$MCS.

\begin{figure}[h]
\begin{center}
\includegraphics[height=6.8cm]{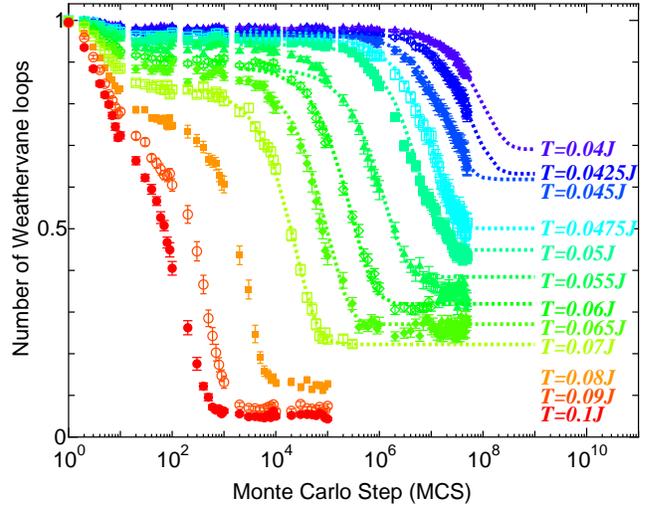}
\end{center}
\caption{(Color online) Relaxation of $n_{\mathrm{loop}}$ from 
$\sqrt{3} \times \sqrt{3}$ structure at several temperatures.
Dashed lines denote the fittling curves estimated by Eq.~(2).}
\label{Fig:relaxsqrt3}
\end{figure}

In Fig.~\ref{Fig:relaxsqrt3}, we plot the relaxation process of $n_{\mathrm{loop}}$ in the case (a) at various temperatures below the critical temperature. 
There, we find two steps in relaxation. 
The first relaxation corresponds to initial local relaxation from the complete ground state configuration. The relaxation time of this process is very short and does not strongly depend on the temperature.
In the second relaxation, the re-construction of the WLs takes place.
The first and second relaxations correspond to energetical and entropical relaxations, respectively.
It is noted that the two-step relaxation is a characteristics of the ordered state
of the present model.

We fit all the second relaxation processes in the following form
\begin{equation}
 n_{\mathrm{loop}} \left( t \right) = A\left( T \right) 
 \exp \left( -\frac{t}{\tau \left( T \right)} \right) 
+ n_{\mathrm{loop}}^{\mathrm{\left( eq \right)}}\left( T\right) ,
\label{nloopt}
\end{equation}

\begin{figure}[h]
\begin{center}
\includegraphics[height=6cm]{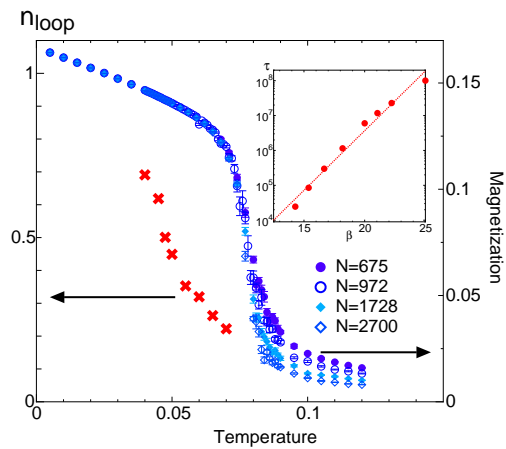}
\end{center}
\caption{
(Color online)
Temperature dependencies of equilibrium values of 
magnetization (closed circles for $N=675$, open circles for $N=972$,
closed diamonds for $N=1728$, open diamonds for $N=2700$, respectively). 
Temperature dependence of equilibrium values of $n_{\mathrm{loop}}$ estimated by Eq.~(2) (crosses).
The inset shows the relaxation time $\tau$ as a function of inverse temperature $\beta$.
The dashed line denotes $\tau=0.53 \mathrm{e}^{0.78\beta}$.
}
\label{Fig:eqloop}
\end{figure}

$\tau \left( T \right)$ denotes a characteristic relaxation time of the 
$n_{\mathrm{loop}}$ and $n_{\mathrm{loop}}^{\mathrm{\left( eq \right)}}$ 
is the equilibrium value of $n_{\mathrm{loop}}$.
Because the first relaxation is much faster than the second one,
the choise of start time of the second relaxation is irrelevant.
Here we fit the curves in the cases of $T=0.07J$, $0.065J$, $0.06J$, $0.055J$, $0.05J$, $0.0475J$, $0.045J$, $0.0425J$ and $0.04J$. 
The temperature dependence of $\tau$ and $n_{\mathrm{loop}}^{\mathrm{\left( eq \right)}}$ are plotted in Figs.~\ref{Fig:eqloop}.
In Fig.~\ref{Fig:eqloop}(a), we find that $n_{\mathrm{loop}}$ increases when the temperature is lowered. We expect that this value continues to the ground state value $n_{\mathrm{loop}}=1$. 
It should be noted that this quantity is not zero at the critical temperature, and it can not be an order parameter although it is a good indicator of the degree of the order of the structure.
In Fig.~\ref{Fig:eqloop}(b), we find a good linearity, which indicates the 
Arrhenius law $\tau(T)\propto e^{\beta \Delta E}$.
From the slope in Fig.~\ref{Fig:eqloop}(b), the energy barrier $\Delta E$ is estimated about $0.78J \sim \mathcal{O}\left(J \right)$. 
It is expected that this energy is necessary when the WLs are reconnected locally.

We study slow relaxation of spin configuration in the magnetically ordered phase of the Ising-like Heisenberg \kagome antiferromagnets. 
From the view point of the entropy, states with the larger $n_{\mathrm{loop}}$ are preferable, although macroscopic configurations are energetically degenerate and they have the same magnetization.
It is expected that the spin structure becomes $\sqrt{3} \times \sqrt{3}$ structure at $T \to 0+$ limit.

We study the relaxation of the number of the weathervane loops which characterizes the spin configuration, and we found the relaxation obeys the Arrhenius law $\tau \propto \mathrm{e}^{\beta \Delta E}$ with the characteristic energy barrier $\Delta E \sim \mathcal{O}\left( J \right)$.

In the present work, we studied relaxation processes in the system of $N=675$,
and pointed out two different relaxation processes exist.
They are energetical and entropical relaxations, respectively.
However, we do not studied the size-dependence of the relaxation time.
The energetical relaxation is a standard one of the ferromagnetic model where we expect the $\sqrt{t}$
scaling, {\it i.e.} $\tau_{\mathrm{mag}} \sim N \left( = L^2 \right) $.
On the other hand, the latter relaxation time depends on how the reconnection of WLs
take place, and at this moment we do not have definited conclusion yet, which will be studied in near future.

Finally, we would like notify that the present phenomena are the effect of residual entropy, but not the entropy of the excited levels which plays important roles in the reentrant phase transition\cite{Kitatani}.
It is either the entropy effect due to the simple harmonic potential which often plays important role to determine the ground state configuration from continuously degenerate ground state\cite{Kawamura2,Kawamura3}. 


We would like to express our thanks to E.~Vincent for valuable discussions and one of the authors (ST) thanks N.~Todoroki and H.~Katsura for useful discussions. 
This work was partially supported by Grant-in-Aid for Scientific
Research on Priority Areas
"Physics of new quantum phases in superclean materials" (Grant No.
17071011) from MEXT,
and also by the Next Generation Super Computer Project, Nanoscience
Program from MEXT,
and also the 21st Century COE Program at University of Tokyo,
``Quantum Extreme Systems and Their Symmetries'' from MEXT.
The authors thank the Supercomputer Center, Institute for Solid State
Physics, University of Tokyo for the use of the facilities.

\end{document}